\documentclass[aps,prl,reprint,amsmath,amssymb,superscriptaddress]{revtex4-2}

\usepackage[usenames,dvipsnames]{xcolor}
\usepackage{amsmath,amssymb}
\usepackage{graphicx}
\usepackage{hyperref}
\usepackage{braket}
\usepackage[normalem]{ulem}
\usepackage{enumerate}
\usepackage{MnSymbol}
\usepackage{esint}
\usepackage{comment}
\newcommand{\Cite}[1]{\mbox{\cite{#1}}}

\newcommand{\be}{\begin{equation}}
\newcommand{\ee}{\end{equation}}
\def\bes{\begin{subequations}}
\def\esu{\end{subequations}}

\newcommand{\dd}{{\rm d}}

\newcommand{\gone}{g_{\text{1D}}}
\graphicspath{{Figures/}}

\begin{document}

\newcommand{\titleinfo}{Realization of fractional Fermi seas}
\title{\titleinfo}

\author{Yi Zeng} 
\thanks{These authors contributed equally to this work. yi.zeng@uibk.ac.at}
\affiliation{Institut f{\"u}r Experimentalphysik und Zentrum f{\"u}r Quantenphysik, Universit{\"a}t Innsbruck, Technikerstra{\ss}e 25, Innsbruck, 6020, Austria}

\author{Alvise Bastianello}
\thanks{These authors contributed equally to this work. yi.zeng@uibk.ac.at}
\affiliation{CEREMADE, CNRS, Universit\'e Paris-Dauphine, Universit\'e PSL, 75016 Paris, France}

\author{Sudipta Dhar}
\affiliation{Institut f{\"u}r Experimentalphysik und Zentrum f{\"u}r Quantenphysik, Universit{\"a}t Innsbruck, Technikerstra{\ss}e 25, Innsbruck, 6020, Austria}

\author{Zekui Wang}
\affiliation{Institut f{\"u}r Experimentalphysik und Zentrum f{\"u}r Quantenphysik, Universit{\"a}t Innsbruck, Technikerstra{\ss}e 25, Innsbruck, 6020, Austria}
\affiliation{
State Key Laboratory of Quantum Optics Technologies and Devices, Institute of Opto-Electronics, Shanxi University, Taiyuan 030006, P. R. China}

\author{Xudong Yu}
\affiliation{Institut f{\"u}r Experimentalphysik und Zentrum f{\"u}r Quantenphysik, Universit{\"a}t Innsbruck, Technikerstra{\ss}e 25, Innsbruck, 6020, Austria}

\author{Milena Horvath}
\affiliation{Institut f{\"u}r Experimentalphysik und Zentrum f{\"u}r Quantenphysik, Universit{\"a}t Innsbruck, Technikerstra{\ss}e 25, Innsbruck, 6020, Austria}

\author{Grigori E. Astrakharchik}
\affiliation{Departament de Física, Campus Nord B4-B5, Universitat Politècnica de Catalunya, E-08034 Barcelona, Spain}

\author{Yanliang Guo}
\email{yanliang.guo@uibk.ac.at}
\affiliation{Key Laboratory of Quantum State Construction and Manipulation (Ministry of Education), School of Physics, Renmin University of China, Beijing 100872, China}
\affiliation{Institut f{\"u}r Experimentalphysik und Zentrum f{\"u}r Quantenphysik, Universit{\"a}t Innsbruck, Technikerstra{\ss}e 25, Innsbruck, 6020, Austria}

\author{Hanns-Christoph  N{\"a}gerl}\email{christoph.naegerl@uibk.ac.at}
\affiliation{Institut f{\"u}r Experimentalphysik und Zentrum f{\"u}r Quantenphysik, Universit{\"a}t Innsbruck, Technikerstra{\ss}e 25, Innsbruck, 6020, Austria}

\author{Manuele Landini}
\affiliation{Institut f{\"u}r Experimentalphysik und Zentrum f{\"u}r Quantenphysik, Universit{\"a}t Innsbruck, Technikerstra{\ss}e 25, Innsbruck, 6020, Austria}

\begin{abstract}
The Pauli exclusion principle is a cornerstone of quantum physics: it governs the structure of matter. Extensions of this principle, such as Haldane's generalized exclusion statistics, predict the existence of exotic quantum states characterized by fractional Fermi seas (FFS), i.e. momentum distributions with uniform but fractional occupancies. Here, we report the experimental realization of fractional Fermi seas in an excited one-dimensional Bose gas prepared through ramping cycles in the interaction strength. The resulting excited yet stable Bose-gas states exhibit Friedel oscillations, smoking-gun signatures of the underlying FFS. The stabilization of these states offers an opportunity to deepen our understanding of quantum thermodynamics in the presence of exotic statistics and paves the way for applications in quantum information and sensing.

\end{abstract}

\maketitle

Quantum statistics determines the behavior of matter at sufficiently low temperatures. Bosons may condense into a single macroscopic wavefunction, while fermions fill all states up to the Fermi energy, forming a Fermi sea \cite{Bose1924,Fermi1926,Proukakis2025,phillips2012advanced,Giorgini2008}. This simple yet profound distinction underlies phenomena as diverse as superconductivity and superfluidity on the one hand, and metallic conductivity, neutron-star stability, and fractional quantum Hall response on the other. However, in one dimension, statistics and interactions are inherently linked. Exchanging particles necessarily involves scattering, and bosons can acquire fermionic traits~\cite{Wilson2020}. Tomonaga-Luttinger liquid theory captures this universal behavior, predicting that interacting bosons may exhibit emergent Fermi seas characterized by power-law correlations and Friedel oscillations (FO)~\cite{Haldane1981,giamarchi2003quantum,Delft1998,Friedel1952}.

It is possible to generalize quantum statistics beyond the dichotomy of bosons and fermions. Inspired by the work on anyons~\cite{PhysRevLett.49.957}, Haldane pioneered generalized exclusion statistics (GES)~\cite{Haldane1991}, where the addition of a particle to the system reduces the available number of states by $\alpha$, interpolating between bosons $\alpha\!=\!0$ and fermions $\alpha\!=\!1$. 
The celebrated anyonic statistics is described by GES with $\alpha \!<\! 1$~\cite{Wu1994}.
Extending the GES idea, one could imagine fractional Fermi seas (FFS) as many-body systems in which each particle occupies an $\alpha\!>\!1$ number of states. One might call the resulting statistics ``super-fermionic". 

Here we report the experimental realization of FFS. Using ultracold one-dimensional (1D) cesium (Cs) gases, we prepare FFS states and measure their momentum distribution for $\alpha\!=\!2$ and $4$. In the first-order correlation functions, we observe Friedel oscillations, enhanced by the non-equilibrium nature of the engineered FFS. We model the experiment using a hydrodynamic approach specifically tailored to nearly integrable models known as generalized hydrodynamics (GHD)~\cite{ThHolo,Alvaredo2016,Bertini2016,Bastianello2022,Doyon2024,Schemmer2019,Moller2021,Malvania2021,Cataldini2022,Schuttelkopf2024,dubois2024,Yang2024Phantom,horvath2025}.

\begin{figure}[b!]
\includegraphics[width=0.85\columnwidth]{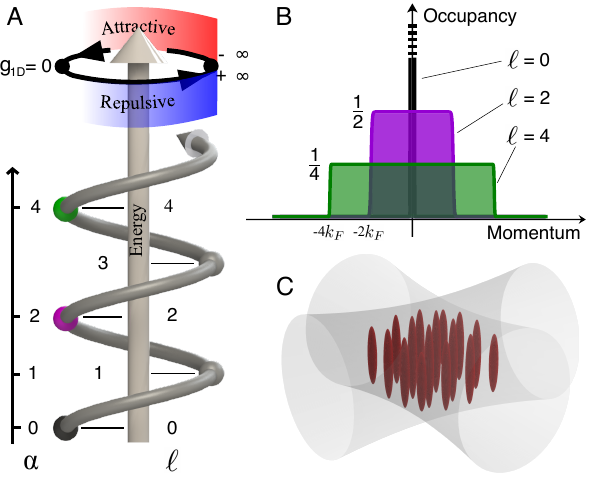}
\caption{\textbf{Emergence of FFS from holonomy cycles.---}
(A): Holonomy cycles are realized by slowly changing the contact interaction from repulsive $\gone\!>\!0$ to attractive $\gone\!<\!0$, passing through the $\gone\!=\!\pm \infty$ point, before again reaching finite repulsive interactions after crossing the non-interacting point $\gone\!=\!0$. The special points on the cycles are indexed by the charge parameter $\ell$ \cite{Marciniak2025,ThHolo}. (B) Momentum-space representation of the FFS for an idealized homogeneous 1D Bose gas. The FFS are realized at the non-interacting points. (C) Experimental platform: Array of vertically oriented 1D tubes (red) generated by a 2D optical lattice (gray shading) filled with Cs atoms.}
    \label{fig:fig1}
\end{figure}

We base our description of the system on the celebrated Lieb-Liniger (LL) model~\cite{Lieb1963,Olshanii1998,Bergeman2003}. It describes interacting bosons in 1D with contact interaction parametrized by the coupling constant $g_{\text{1D}}$. In our cold-atom experiment, $g_{\text{1D}}$ can be freely and dynamically tuned by means of Feshbach resonances in conjunction with a confinement-induced resonance (CIR) ~\cite{Haller2009,Bergeman2003} to values from zero to $+\infty$ and $-\infty$. The coupling constant is given by $g_{\text{1D}}\!=\!2\hbar^2 a_{\text{3D}}/[ma^2_\perp(1-1.0326a_{\text{3D}}/a_\perp)]$~\cite{Olshanii1998}, where $m$ is the atoms' mass and $a_\perp$ is the transverse harmonic oscillator length $a_\perp=\sqrt{\hbar/m\omega_\perp}$ of our 1D trap with transverse frequency $\omega_\perp$, and $a_{\text{3D}}\!=\!a_{\text{3D}}(B)$ is the 3D scattering length, which we tune by means of a magnetic field $B$. FFS have recently been predicted to exist for the LL model as excited many-body states~\cite{Marciniak2025,ThHolo}. It was proposed that repeated holonomy cycles~\cite{Yonezawa2013} would give access to a succession of excited states. 
In these cycles~\cite{Kao2021,Marciniak2025} the coupling $\gone$ is ramped from the non-interacting value $\gone\!=\!0$ to the strongly repulsive regime ($\gone\!=\!+\infty$), where the system realizes the Tonks-Girardeau (TG) phase~\cite{Kinoshita2004,paredes2004,Batchelor2007}. Then, the coupling is switched to strongly attractive values ($\gone\!=\!-\infty$), where the gas is in the super-Tonks-Giradeau (sTG) state~\cite{Haller2009,Kao2021,Batchelor2005}. Finally, $\gone$ is ramped back to $\gone\!=\!0$. The state of the system is parameterized by the charge parameter $\ell$,  as shown in Fig.~\ref{fig:fig1} (A). 
As we shall see in the following, due to the collective 
nature of our quantum system, our experiment is better understood in terms of taking the system into thermodynamic states of progressively higher energy, described by generalized Gibbs ensembles (GGE)~\cite{Yang1969,Rigol2007,Langen2015}. In view of this, we refer to our experimental protocol as interaction cycles. The resulting quasi-adiabatic dynamics is expected to induce the formation of the FFS at $\ell\!=\!2,4,...$. As depicted in Fig.~\ref{fig:fig1} (B), the initial delta-function-like momentum distribution for $\ell\!=\!0$ transmutes into a box-type distribution $n(k)\!=\!1/2$ for $|k|\!<\!2k_\text{F}$ for $\ell\!=\!2$ and then to $n(k)\!=\!1/4$ for $|k|\!<\!4k_\text{F}$ for $\ell\!=\!4$. Here, $k_\text{F}$ is the Fermi wave-vector of the TG state. With this parameterization, we can identify $\alpha\!=\!\ell$. 
Even though the FFS are highly excited many-body states, they are stabilized by the integrability of the LL model~\Cite{Marciniak2025,ThHolo}. Note that the FFS are realized at the non-interacting points. 

Our experiment begins by preparing a pure Cs Bose-Einstein condensate (BEC) in the Thomas-Fermi regime with no detectable thermal fraction and an atom number of $5\!\times\! 10^4$ with $\pm5\%$ shot-to-shot fluctuations~\cite{Sup}. The scattering length is initially set to $a_{\text{3D}}\!=\!195(1) \, a_0$. Here, $a_0$ is Bohr's radius. We adiabatically load the BEC into a 2D optical lattice with a spacing of 532.35~nm and depth of 30 $E_\text{r}$, where $E_\text{r}\!=\!\hbar^2k^2_\text{L}/(2m)$ is the photon recoil energy with the lattice wavevector $k_\text{L}$. As part of this procedure, $a_{\text{3D}}$ is ramped to $a_{\text{3D}}\!=\!517(1)a_0$ during lattice loading. The atoms then populate a non-uniform array of about $7000$ 1D tubes (Fig.~\ref{fig:fig1}(C)) with tight transverse confinement ($\omega_{\perp}/2\pi\!=\!15(1)~\text{kHz}$) and weak longitudinal harmonic confinement ($\omega_{||}/2\pi\!=\!18(1)~\text{Hz}$) \cite{Guo2024} with an average population of 10 atoms per tube and an average 1D density of $n_{\text{1D}}\!\simeq\!0.5/\mu$m. For implementing the interaction cycles, $g_{\text{1D}}$ is tuned in the vicinity of a Feshbach resonance with its pole at $B\!=\!47.78(1)$~G \cite{Haller2009}. Given the transverse confinement, the pole of the CIR is at $B\!=\!47.27(1)$~G.

\begin{figure}[t]
    \centering
    \includegraphics[width=\columnwidth]{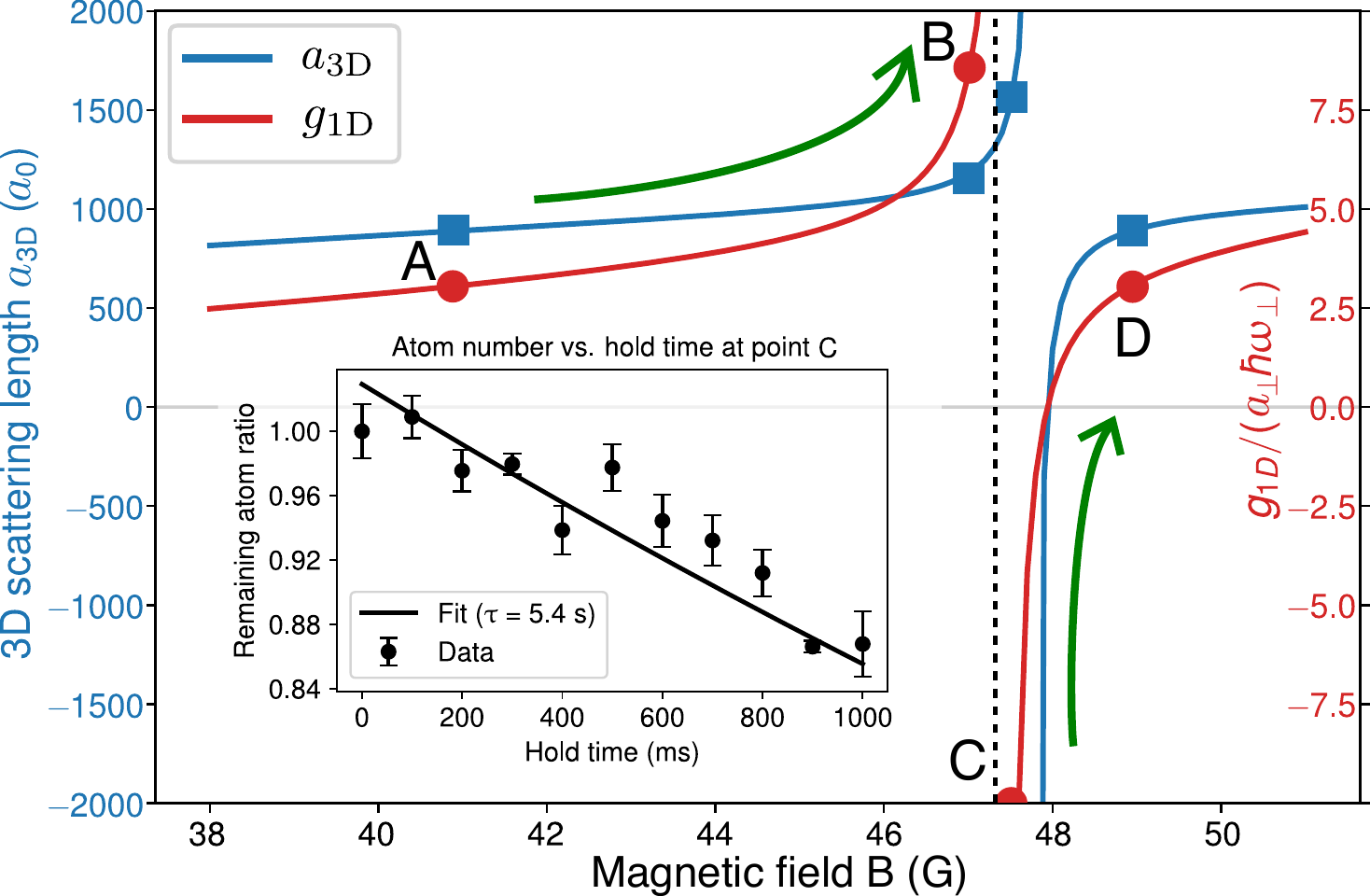}
    \caption{\textbf{The implementation of interaction cycles---} We plot $a_{\text{3D}}$ and $g_{\text{1D}}$ as a function of magnetic field strength $B$. One cycle consists of one $B$-field ramp, one jump, another $B$-field ramp, and one quench between the field values labeled as A to D. The green arrows indicate slow ramps from point A to B and point C to D. The jump from point B to C and the quench from point D to A close out the cycle. The $B$-field values of these points are $B_\text{A}\!=\!40.85$~G, $B_\text{B}\!=\!47.20$~G, $B_\text{C}\!=\!47.64$~G and $B_\text{D}\!=\!49.01$~G. The vertical dashed line indicates the CIR's pole. The inset shows an atom-loss measurement for the sTG state at point C that gives a $1/e$-lifetime of $5.4$ s.}
    \label{fig:fig2}
\end{figure}

Figure~\ref{fig:fig2} illustrates how the interaction cycles are performed experimentally. The protocol is optimized to minimize the excitation of collective motion in the trap. Initially, for $a_{\text{3D}}\!=\!517(1) a_0$ at $B\!=\!27.90(1)$~G, the interaction strength is $\gamma\!\simeq\!34$, where $\gamma\!=\!m g_{\text{1D}}/(\hbar^2n_{\text{1D}})$ is the Lieb-Liniger parameter. It is then ramped to the starting point of the cycle at $B_\text{A}$ in $500$~ms, where the system is strongly repulsive with $\gamma\!\simeq\!120$ deep in the TG regime. The first step of the cycle goes from $B_\text{A}$ to $B_\text{B}$ in 200~ms, approaching the CIR's pole. The next step is a small jump across the CIR from $B_\text{B}$ to $B_\text{C}$ in $\sim0.3$~ms. The system is now in the sTG regime, with $\gamma\!\simeq\!-134$. This is then followed by a slow ramp to $B_\text{D}\!=\!49.01$~G in 250~ms. Note that this ramp includes a zero crossing $g_{\text{1D}}\!=\!0$ that is traversed from the negative to the positive side. At $B_\text{D}$ the interaction strength is the same as at $B_\text{A}$. The last step in the cycle $D\rightarrow A$ is a fast jump within $\sim0.2$~ms back to the initial point on the other side of the CIR. Using this scheme, we can perform multiple cycles on the same CIR.

\begin{figure*}[!t]
    \centering
\includegraphics[width=2\columnwidth]{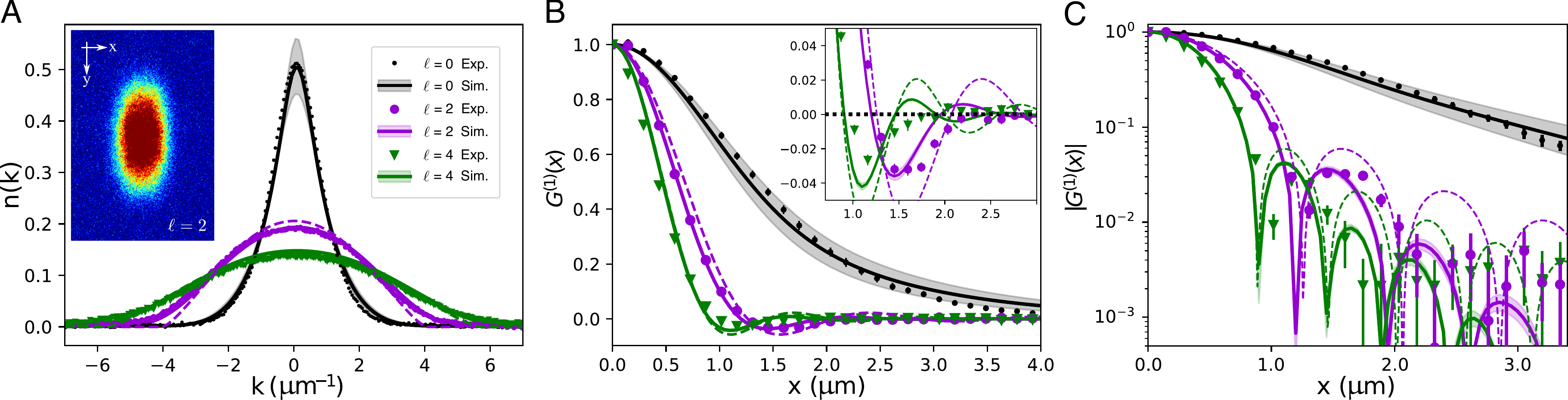}
    \caption{\textbf{Evidence for FFS.---} (A) Momentum distributions $n(k)$ of the ideal Bose gas for $\ell=\!=\!0, 2,$ and $4$, produced by integrating the ToF absorption images (inset) over the transversal $y$ direction, comparing the experimental data (solid points) to the GHD simulation data (solid lines) and the analytical prediction using Thomas-Fermi approximation for $\ell\!=\!2$ and $\ell\!=\!4$ (dashed lines). The shaded regions show the effect of thermal fluctuations on the GHD simulation, by changing the fitted 1D temperature $T_\text{1D}=\!=\!5\text{ nK}$ by $\pm5\%$. First-order correlation function $G^{(1)}(x)$, in linear scale (B) and for its absolute value in log scale in (C). The inset in (B) is a zoom-in around the zero crossing. All experimental data is the average of 10 repetitions, with error bars reflecting the standard error of the mean. When not visible, the error bars are smaller than the symbols.} 
    \label{fig:fig3}
\end{figure*}

Implementing the interaction cycle in the experiment poses a significant challenge, as it requires slowly traversing the unstable attractive regime~\cite{Astrakharchik2005,Gerton2000,Donley2001}. In theory, the integrable nature of the LL model protects the system from collapsing. Integrable systems possess an extensive number of conserved quantities that strongly constrain their evolution, allowing them to stabilize non-thermal states well described by GGEs~\cite{Yang1969,Rigol2007, Langen2015}. 
In practice, the inevitable presence of small integrability-breaking perturbations in realistic experimental settings, primarily from three-body collisions, limits the stability of the system in the attractive regime~\cite{Haller2009}. Previous works introduced extra integrability-breaking dipolar interaction to mitigate the issue~\cite{Kao2021,Chen2023}. Our loading procedure at $a_{\text{3D}}\!=\!500a_{0}$ helps in this regard: strong repulsive interactions lead to low-density, yet strongly interacting 1D gases. This stability on the repulsive side of the resonance is continued to the attractive side. As shown in Fig.~\ref{fig:fig2}(inset), the gas in the sTG regime has a lifetime of more than $5$s.

We are able to implement two and a half cycles without significant atom loss. To characterize the excited ideal-gas states corresponding to the points $\ell\!=\!2$ and $\ell\!=\!4$ in the cycle, we slowly ramp the magnetic field $B$ within $500$~ms from point~A of the corresponding cycle to $B\!=\!17.15$~G, where $\gone\!=\!0$~\cite{Gustavsson2008} and the slope $\dd\gone/\dd B$ is very small, about a factor of $110$ less than that of the zero-crossing point near the CIR, between points~C and D. This facilitates the subsequent measurement in view of ambient magnetic-field noise. We measure the momentum distribution $n(k)$ via time-of-flight (ToF) expansion. The measurement for $\ell\!=\!0$
is taken before the first cycle, by ramping $B$ directly to $B\!=\!17.15$~G in $200$~ms. The measurement ramp includes a fast but low-amplitude jump over a narrow Feshbach resonance at $B\!=\!19.8$~G \cite{Mark2007}.

Figure~\ref{fig:fig3}(A) reports the momentum distributions and their comparison with analytical predictions and numerical simulation results based on GHD. The $\ell\!=\!0$ state exhibits a comparatively narrow distribution with FWHM width of $1.72~\mu\text{m}^{-1}$, as expected for a low-energy state of non-interacting bosons broadened by a finite temperature. Upon reaching the $\ell\!=\!2$ and $\ell\!=\!4$ points, the distributions broaden and flatten significantly, with FWHM widths of 5.19 and 6.69~$\mu\text{m}^{-1}$. In a homogeneous system, these excited states would exhibit the FFS profile as sketched in Fig.~\ref{fig:fig1}(B)~\cite{Marciniak2025,ThHolo}; in the experiment, harmonic confinement and averaging over the non-uniform tube distribution smooth out its sharp features. 

The first-order correlation functions averaged over the tubes $G^{(1)}(x)$~\cite{Guo2024,Guo2024cross} are shown in Fig.~\ref{fig:fig3}(B,C). These are obtained by Fourier transforming the measured momentum distribution~\cite{Bloch2008,Hove1954}. Here, $x$ is the direction along the 1D tubes. We observe that for the $\ell\!=\!0$ state, $G^{(1)}(x)$ remains positive and decays smoothly, as expected for a finite-temperature bosonic gas. For the excited states $\ell\!=\!2$ and $\ell\!=\!4$ the coherence decays more rapidly. This is expected in view of the broader momentum distributions for higher $\ell$. Remarkably, the correlation functions for $\ell\!=\!2$ and $\ell\!=\!4$ develop an oscillatory behavior. Indeed, $G^{(1)}(x)$ dips below zero, showing pronounced minima at approximately $1.6~\mu\text{m}$ for $\ell\!=\!2$ and $1.2~\mu\text{m}$ for $\ell\!=\!4$. This behavior reflects the onset of Friedel oscillations (FO), and provides clear evidence for the formation of FFS. For a homogeneous FFS one would expect a $G^{(1)}(x)$ that is a cardinal sine function, the well-known Fourier transform of the single slit from elementary optics, for which the stretched Fermi momentum determines the FO period~\cite{Marciniak2025,ThHolo}. This simple expression can be modified to account for the trapping potential and tube averaging~\cite{Sup}. Specifically, we employ the Thomas-Fermi approximation as appropriate for fermionic systems, and approximate the population of atoms in the tubes as a flat distribution for $N\!\le\!N_\text{max}$. We get a momentum occupancy
\begin{equation}
n(k)\simeq\frac{2 a_\parallel}{3\pi  \ell^2 N_\text{max}^2}[2\ell N_\text{max}-a_\parallel^2 k^2]^{3/2}
\end{equation}
and with this 
\begin{equation}
    G^{(1)}(x)=\frac{8}{(x a_\parallel^{-1})^2 2 \ell N_\text{max}}J_2(x a_\parallel^{-1} \sqrt{2 \ell N_\text{max}}),
\end{equation}
where $J_2$ is the second-order Bessel function and $a_\parallel\!=\!\sqrt{\hbar/(m\omega_\parallel)}$ is the oscillator length of the harmonic trap along the longitudinal direction. The continuous parameter $N_\text{max}$ is determined by normalization within the Thomas-Fermi approximation~\cite{Sup}. For our experiment, we get $N_\text{max}=17.45$. 
These simple formulas already capture well the experimental FO data shown in Fig.~\ref{fig:fig3}. We attribute the corrections predominantly to thermal fluctuations and the deviation from a homogeneous tube occupancy. Both effects become less relevant upon increasing $\ell$, and they are included in the GHD simulations, which give a better agreement with our data.

The GHD simulation accounts for the trapping potential \cite{Doyon2017}, the time-dependent interactions \cite{Bastianello2019}, and finite-temperature effects in the experimental protocol (see SM~\cite{Sup}). It has two free parameters: the 1D temperature $T_\text{1D}$ (assumed homogeneous across the tubes) and a dimensional cross-over temperature $T_\text{co}$, which governs the atom-number distribution at the transition from the 3D BEC to 1D tubes \cite{horvath2025}. We calibrate these parameters by fitting the $\ell\!=\!0$ momentum distribution, obtaining $T_\text{1D}\simeq 5\, \text{nK}$ and $T_\text{co}\simeq 2 \, \text{nK}$; the $\ell\!=\!2,4$ distributions then follow directly from the simulated dynamics with slow ramps in the long-time limit. The shaded regions in Fig.~\ref{fig:fig3} indicate simulations with small variations of these fitted parameters~\cite{Sup}: their effect is mainly visible for $\ell\!=\!0$, while the excited-state momentum distributions at $\ell\!=\!2$ and $\ell\!=\!4$ are robust. Evidently, GHD reproduces our data well. Note that both the numerical and analytical models predict further oscillations in $G^{(1)}(x)$ beyond the first two nodes. These are, however, masked by the noise floor. 

\begin{figure}[t]
    \centering \includegraphics[width=\columnwidth]{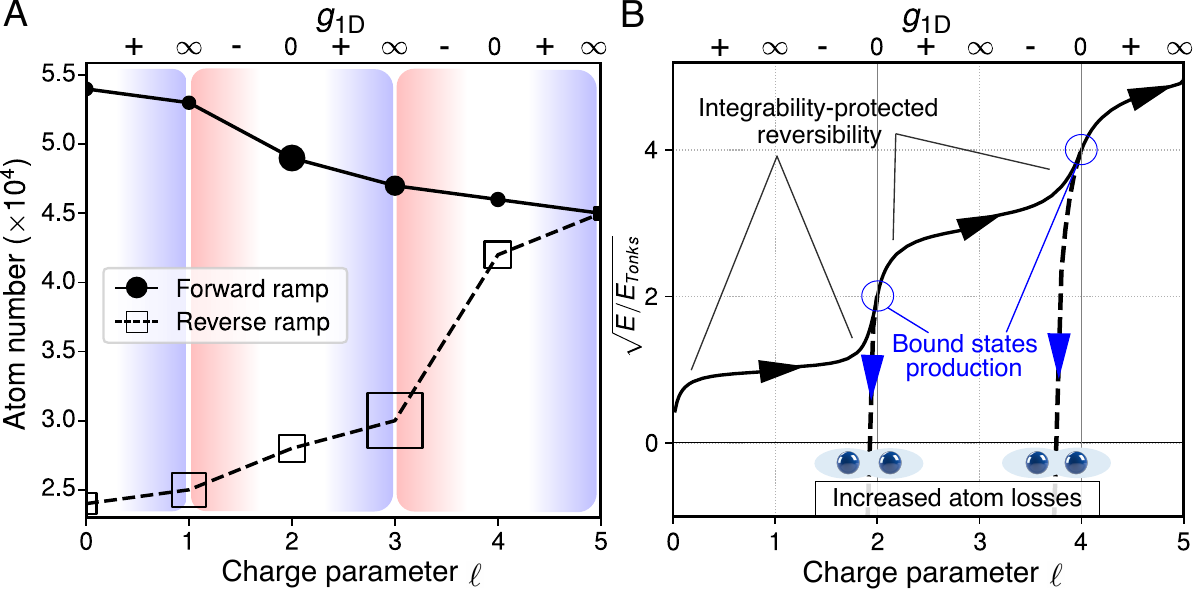}
    \caption{\textbf{Irreversibility of the interaction cycles.---} (A) Measurement of atomic losses in forward and backward ramping. The remaining atom number is measured on each step of the forward and reverse cycles. Red shading indicates attractive interaction and blue shading indicates repulsive interaction. The size of the marker is proportional to the fractional atom loss from the previous step of the ramp. (B) Energy evolution during the cycles, calculated from the GHD simulation. The solid line indicates the total energy of the system versus $\gone$ in the forward ramp of the interaction cycles. The dashed lines give the energy evolution when ramped in reverse from the repulsive excited states. The overall energy decreases sharply due to the negative binding energy.
    }
    \label{fig:fig5}
\end{figure}

For any type of cycle it is natural to ask whether the process is reversible. We therefore, after having crossed the point $\ell\!=\!4$ and having reached the next TG regime at $\ell\!=\!5$, run the protocol backward. Figure~\ref{fig:fig5}(A) shows the atom number during the forward and reverse ramp: while in the forward protocol atom losses are limited to about $20\%$, the reverse cycle loses up to $45\%$ of the atoms. This signals a strong asymmetry depending on the direction in which the cycle is performed. In the inverse protocol, losses are concentrated around the first passage through the zero crossing, towards attractive interaction. This evident irreversibility is consistent with the GHD predictions ~\cite{Koch2021,ThHolo} on accessing the attractive regime with slow ramps through the non-interacting point. In this case, bound states, so-called Bethe strings, are excited even when the dynamics is integrable~\cite{horvath2025}. Their formation significantly increases atom loss when the interaction becomes more attractive, and bound states become increasingly more unstable as their binding energy grows, see Fig.~\ref{fig:fig5}(B). 
The atom losses observed here are compatible with the expectation that most of the atoms that are able to form bound states are lost already in the first reverse ramp between $\ell\!=\!4$ and $\ell\!=\!3$. After that, further bound states are less likely to be excited and losses are more contained, probably due to the reduced density. Note that in the idealized scenario of holonomy cycles of the interaction strength~\cite{Yonezawa2013,Marciniak2025} the gas accesses excited states through a quantum adiabatic operation, thus following a single eigenstate of the instantaneous Hamiltonian. Under this assumption the protocol would be reversible, in contrast to the GHD prediction where the system explores a macrostate characterized by a GGE. Our observations hence support the GHD interpretation.

In conclusion, we prepared long-lived excited states of a 1D Bose gas with FFS momentum distributions, using interaction-ramping cycles. In the associated first-order correlation functions we observe the onset of FOs, providing a direct signature of the engineered Fermi-surface structure. Correlation functions of ultracold bosons in the ground state are described by the Tomonaga-Luttinger liquid theory and FOs are subleading with respect to a smooth power-law decay~\cite{Delft1998}.
In contrast, interaction-ramping cycles realize non-equilibrium states with prominent FOs suggesting an emergent field-theory description beyond the conventional Tomonaga-Luttinger liquid. A dedicated analysis of the long-range correlations is presented in an accompanying theoretical work~\cite{ThHolo}.
Finally, our results establish interaction-ramping cycles as a controllable route to non-equilibrium states in a nearly integrable many-body system, offering an experimental platform to explore generalized exclusion statistics and its dynamical consequences. The associated FOs provide a direct diagnostic of the underlying many-body structure.  An exciting future prospect would be to extract the second-order correlation function of the FFS using impurities as local probe. 
Fermi seas deeply affect the dynamics of injected impurities, with emergent superfluidity~\cite{horvath2026} and Bloch oscillations in the presence of external forces~\cite{Dhar2025,Meinert2017}. It is therefore natural to investigate these phenomena on FFS. More broadly, geometric control of highly excited many-body states may enable new approaches to the engineering of correlations, transport properties, and response in quantum simulators. 

\bigskip
\begin{acknowledgments}
The Innsbruck team acknowledges funding by the European Research Council (ERC) with project number 101201611 and by an FFG infrastructure grant with project number FO999896041. Y.Z. and  Y.G. are supported by the Austrian Academy of Sciences (\"OAW) with APART-MINT 12234 and by the Austrian Science Fund (FWF) with project number 10.55776/COE1, respectively. Z.W. is supported by the Quantum Science and Technology-National Science and Technology Major Project (Grant No. 2021ZD0302003) and the National Natural Science Foundation of China (Grant Nos. 12488301, 12034011, U23A6004). A.B. acknowledges partial support of the German Research Foundation (DFG) under the Germany's Excellence Strategy-EXC-2101-3990814868. G.E.A. acknowledges financial support from Ministerio de Ciencia e Innovación MCIN/AEI/10.13039/501100011033 (Spain) under Grant No. PID2023-147469NB-C21. \textbf{Author Contributions:} This work was conceived by Y.Z., Y.G., H.C.N. and M.L.. Experiments were prepared and and performed by Y.Z., S.D., X.Y., M.H. and Y.G.. Data were analyzed by Y.Z., A,B., Z.W. and Y.G.. Numerical simulations and analytical prediction were performed by A.B. and G.E.A. The manuscript was drafted mainly by Y.Z., A.B., M.L. and H.C.N. All authors contributed to the discussion and finalization of the manuscript. \textbf{Data Availability:} The data shown in the main text are available via Zenodo~\cite{Zenodo}. \textbf{Code Availability:} Codes supporting the findings of this study are available from the corresponding author upon reasonable request.
\end{acknowledgments}

\bibliography{biblio_exp_new}
\end{document}


\beginsupplementA

\begin{figure}[b]
    \centering
    \includegraphics[width=\columnwidth]{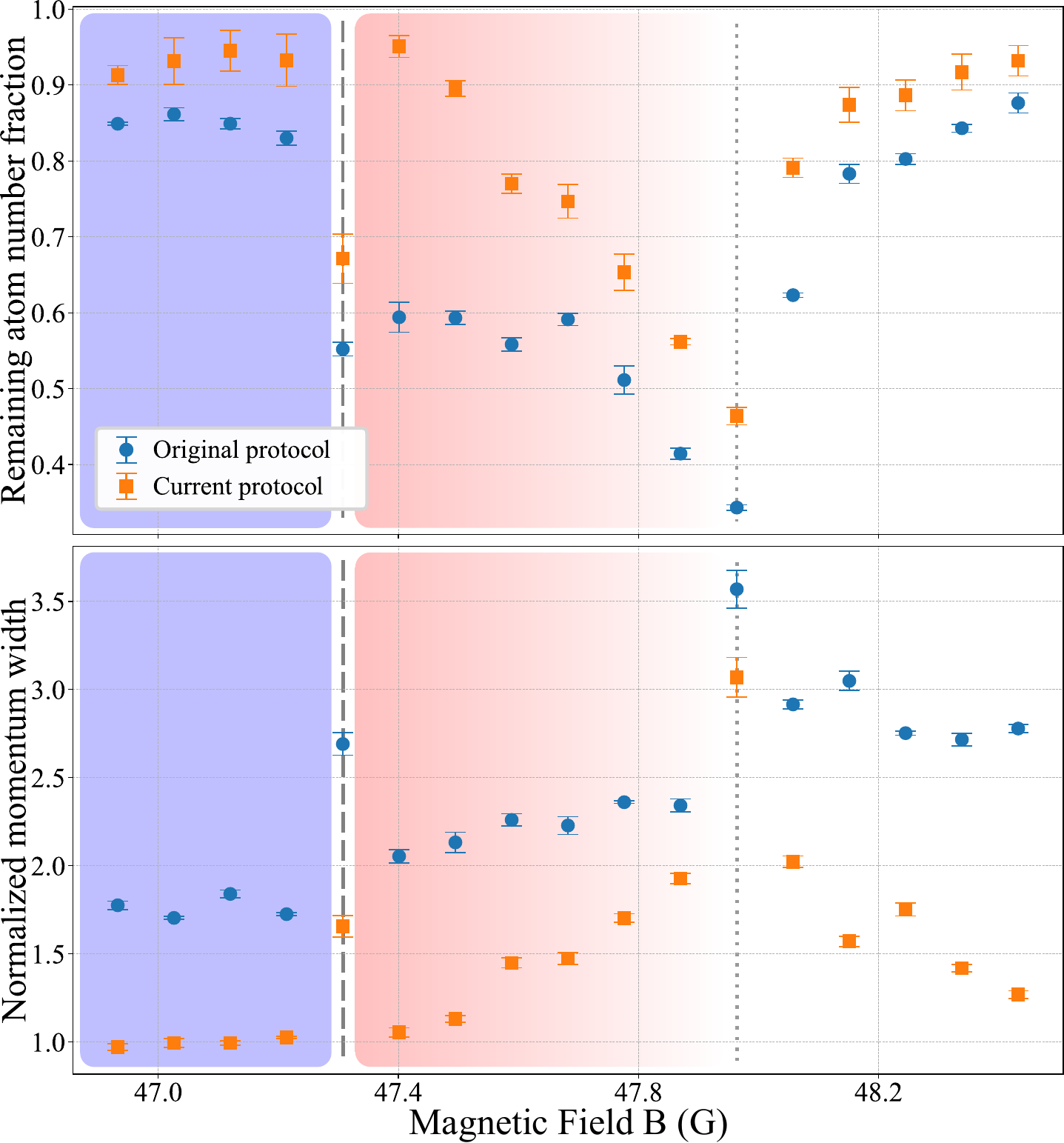}
    \caption{ \textbf{Stability of the sample in the attractive regime.---} Relative atom number (top) and momentum width (bottom) measured after a 200 ms hold time across the CIR. Blue (red) shaded regions indicate the repulsive (attractive) interactions. The dashed and dotted lines indicate the positions of the CIR and of the zero-interaction point, respectively. All experimental data is the average of 3 repetitions, with error bars reflecting the standard error of the mean. When not visible, the error bars are smaller than the symbols. }
    \label{fig:S1}
\end{figure}
The supplementary material provides additional experimental details in Notes 1--4 and details about the theory in Notes 5--7. 

\section{Supplementary note 1: Enhanced Stability of the sTG gas}
\label{sec_stability}

The implementation of the parameter cycles in the experiment relies critically on the stability of the quantum gas throughout the interaction ramping sequence, particularly in the strongly-attractive sTG regime ($\gamma \rightarrow -\infty$). 
In principle, the sTG gas should be stable thanks to integrability, which suppresses decay and clustering. However, integrability-breaking perturbations (e.g., the harmonic trap and three-body processes) can lead to heating and atom loss~\cite{Haller2009}.

We tested different state-preparation protocols to improve the stability of the gas.
We compare our protocol with the one originally employed in the first realization of the TG-sTG transition~\cite{Haller2009}, hereafter we call the original protocol, where the gas was observed to possess limited stability across the whole attractive regime. In Ref.~\cite{Haller2009}, tubes are loaded at $a_\text{3D}\!\approx\!200a_0$ and the quench to the sTG regime is intiated at $a_\text{3D} \!\approx\!900 a_0$. In our protocol, the gas was loaded into the lattice at $a_\text{3D} \!\approx\! 500a_0$, resulting in a significantly lower atomic density. The jump to sTG is initiated at $a_{3D} \!\approx\!  1200 a_0$. 

\begin{figure}[b]
    \centering
    \includegraphics[width=\columnwidth]{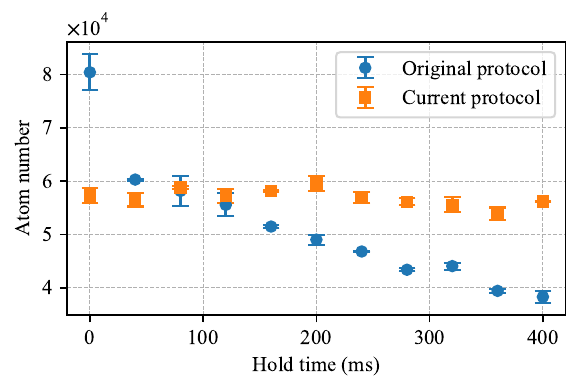}
    \caption{\textbf{Improved lifetime.---} We compare the original and current preparation protocols, plotting the number of remaining atoms versus the hold time in the sTG regime ($B\!=\!47.57$~G). All experimental data is the average of 3 repetitions, with error bars reflecting the standard error of the mean. When not visible, the error bars are smaller than the symbols.}
    \label{fig:S2}
\end{figure}
 
Figure~\ref{fig:S1} shows the relative atom number and momentum width (extracted from ToF expansion) measured after holding the gas for 200 ms at various values for the magnetic field $B$ across the CIR. Measurements are normalized to the corresponding initial values in the TG regime ($B\!=\!47.1$~G). 
With the original protocol, the final position where we hold the $B$-field is reached with a 0.2-ms quench from $B\!=\!42.8$~G, as is done in Ref.~\cite{Haller2009}. We observe significant atom loss and a sharp increase for the momentum width immediately upon entering the sTG regime and near the zero crossing of the interaction strength. This indicates loss and heating associated with bound-state formation. Compared with the results from Ref.~\cite{Haller2009}, the stability is slightly improved, largely due to experimental upgrades (in particular due to improved magnetic-field control and stabilization).

With the protocol we used to realize FFS, the final position where we hold the $B$-field are reached with a 0.2-ms quench from $B\!=\!47.1$~G. The stability is dramatically improved: atom loss is substantially reduced and the momentum-width increase is strongly suppressed across the full range of magnetic-field values.

Figure~\ref{fig:S2} shows a lifetime measurement in the deep sTG regime ($B\!=\!47.57$~G). The system 
exhibits a lifetime exceeding 5 s, confirming the robust stabilization achieved through optimized preparation. Importantly, the comparison in Fig.~\ref{fig:S2} also shows that the improved stability is not solely a consequence of a lower atom number, since the cloud prepared with the original protocol continues to lose atoms at a faster rate even when its atom number becomes smaller than that of the current protocol.
\begin{figure}[t!]
    \centering
    \includegraphics[width=\columnwidth]{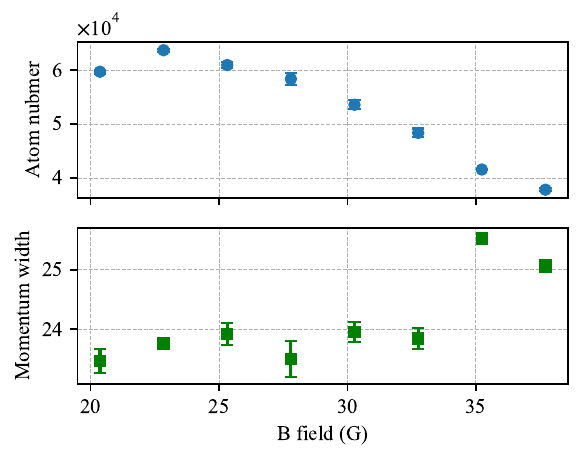}
    \caption{
   \textbf{Measurements of atom loss and heating for different loading conditions.---} We load the 1D tubes at different values of the magnetic field, thus changing the interaction strength during the 1-second loading time. After unloading the lattice adiabatically we measure the remaining atom numbers and the width of the resulting momentum distribution. All experimental data is the average of 3 repetitions, with error bars reflecting the standard error of the mean. When not visible, the error bars are smaller than the symbols.}
    \label{fig:S2.5}
\end{figure}

\section{Supplementary note 2: Optimization of the Loading Conditions}
\label{sec_optimization_loading}

Loading the array of 1D tubes at higher repulsive interactions has the advantage of spreading out the atoms to more tubes, thus lowering the number density within single tubes, without sacrificing the signal-to-noise ratio, unlike starting with a BEC with lower total atom number. 

Here we characterize the heating and losses caused by loading the system into 1D tubes for different values of $a_\text{3D}$. The procedure involves preparing a 3D BEC and adiabatically ramping up the 2D lattice to form the array of 1D tubes, for different values for $B$, corresponding to different values for $a_\text{3D}$. After a brief hold time, we ramp the lattice back down with the same ramp rate. Finally, we measure the atom number and momentum width of the resulting BEC. Figure~\ref{fig:S2.5} shows the results. At higher values for $B$ (i.e, higher interaction strength), we observe significant atom loss and momentum-distribution widening. The atom number and momentum width show only very small variations for $B$-field values between 20 G and 28 G (corresponds to $a_\text{3D}$ between $200a_0$ and $500 a_0$). This indicates that loading the tubes at  $500a_0$ does not introduce significant loss or heating compared to doing this at lower interaction strengths.

\section{Supplementary note 3: Optimization of the Interaction Ramps}
\label{sec_optimization_ramp}

The interaction-strength ramps are optimized by balancing two contrasting effects. Fast ramps prominently excite the breathing mode\cite{Haller2009}, resulting in damped oscillations in the width of the momentum distribution . In contrast, too slow ramps increase atom losses due to the longer stay in the metastable attractive regime.
\begin{figure}[b!]
    \centering
    \includegraphics[width=\columnwidth]{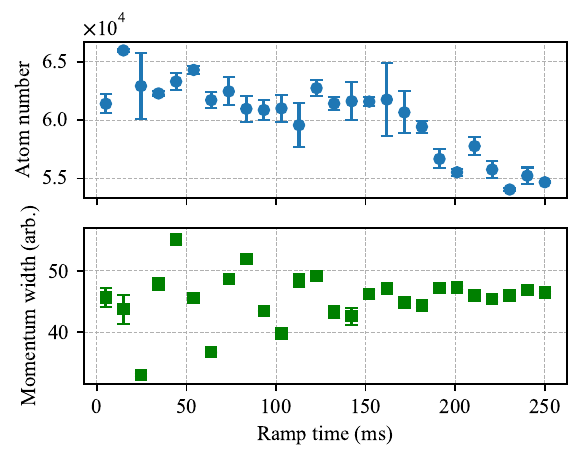}
    \caption{\textbf{Optimization of the ramp duration.} Final atom number (top) and momentum width (bottom) versus the duration of the interaction ramp from the sTG regime across the non-interacting point and into the repulsively interacting regime. Oscillations in the momentum width indicate excitation of the breathing mode \cite{Schemmer2018} for short ramp times. Note that this measurement is done with double the initial atom number of the main experiment, to make the effect more noticeable. For lower atom numbers, the system suffers less atom loss and is less likely to excite the breathing motion. All experimental data is the average of 3 repetitions, with error bars reflecting the standard error of the mean. When not visible, the error bars are smaller than the symbols.}
    \label{fig:S3}
\end{figure}

\begin{figure*}[t!]
    \centering
\includegraphics[width=0.9\textwidth]{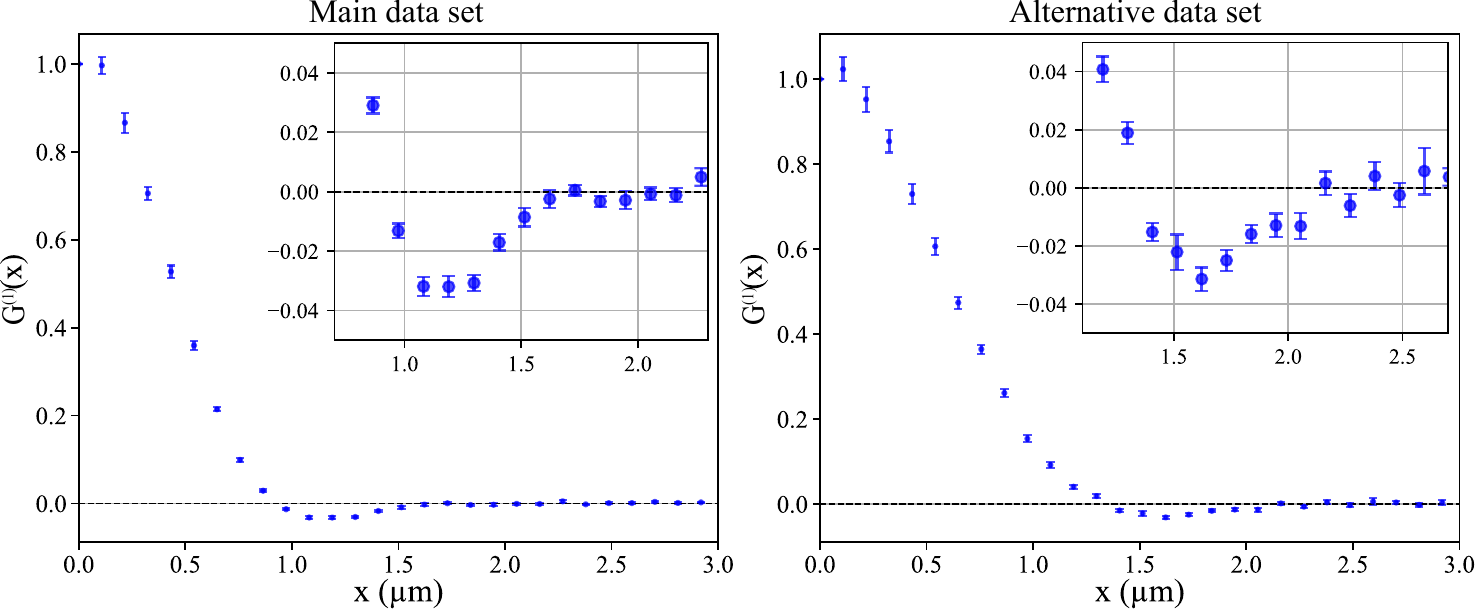}
    \caption{\textbf{Comparison of the FO for different densities.---} We compare $G^{(1)}$ from the main dataset (left) corresponding to $\ell\!=\!2$ with an alternative dataset (right) obtained with different loading conditions. More specifically, alternative data are obtained loading the 1D tubes with approximately half of the total number of atoms used in the main dataset. All the other loading parameters and the interaction cycle protocol remain unchanged.
    The position of the first zero crossing moves to a larger distances with a lower atom number. All experimental data is the average of 10 repetitions, with the error bars reflecting the standard error of the mean.}
    \label{Fig_SM_oldnew}
\end{figure*}

We therefore optimize the ramp duration by measuring the momentum width and the remaining atom number after a full ramp sequence (at point D in Fig.~2(C) of the main text) as a function of the duration of the final ramp (from point C to point D), see Fig.~\ref{fig:S3}.
The momentum width exhibits damped oscillations as a function of the ramp time. These oscillations correspond to the excitation of the longitudinal breathing mode, indicating a non-adiabatic process for short ramp times. The oscillations damp out significantly for ramp times longer than approximately 150 ms.
The number of remaining atoms shows a monotonic decrease with increasing ramp time, with a stronger drop after 150 ms, see Fig.~\ref{fig:S3}(b). We choose a ramp time of 250 ms for the experiments presented in the main text as a compromise between limiting atom loss and not exciting too much the breathing motion.

\section{Supplementary note 4: The robustness of the Friedel oscillation}
\label{sec_Friedel}

In Fig.~\ref{Fig_SM_oldnew} we demonstrate the robustness of the FO by comparing the oscillation as presented in the main text with that of an alternative data set, in the $l$=2 excited ideal-gas state. For this alternative dataset, we reduced the atom number by a factor of two ($N\sim 2.5\times 10^4$).

Although the experimental resolution makes it difficult to appreciate more than a single oscillation, we can qualitatively compare the two data sets. As discussed in the main text, we can analytically describe the excited ideal gas states of our experiment by giving the FFS momentum distribution a density dependency, and using the Thomas-Fermi approximation to estimate the density distributions, see Supplementary Note~\ref{sec_TF}). For states with the same $\ell$, halving the initial atom number should result in approximately halving the number of atoms in the central tube and rescaling the position of the first zero in $G^{(1)}$ by a factor $\sqrt{2}$. This is consistent with the data in Fig.~\ref{Fig_SM_oldnew}. 
Future experimental upgrades, such as the implementation of a box trap, will enable the precise measurement of the FO frequency for a more precise comparison.

\section{Supplementary note 5: Theoretical description of the state preparation}
\label{S_sec_prep}

Our theoretical modeling of the experiment encompasses two steps: first, we determine the initial state experimentally prepared before the interaction cycle protocol, accounting for the 1D inhomogeneity caused by the longitudinal trapping, the number of atoms varying across the tubes, and finite-temperature effects. Then, the so-determined initial state is evolved through the protocol: this second step is performed with GHD and discussed in the \secref{S_sec_GHD}, here we present the modeling of our state-preparation protocol.

To describe the initial state, we follow the same approach as presented in Ref.~\cite{horvath2025}. The gas is initially prepared in a BEC in a 3D trapping potential, then the power of the transverse lattice is ramped up until the energy of the transverse modes is sufficiently high that one enters the 1D regime. We model this process in two steps, for low lattice depths, particles have a significant probability to tunnel between tubes. As the transverse trapping becomes stronger, the inter-tube tunneling time-scale becomes much longer than the experimentally relevant time scales. We refer to this transition as ``dimensional crossover".
Hence, we consider the following approximate description of the state preparation:
\begin{enumerate}
\item In the 3D regime, before the dimensional crossover, we consider the gas arranged in an array of 1D tubes that can exchange energy and particles. The tubes are at thermal equilibrium, characterized by a global temperature and a chemical potential.
\item At the dimensional crossover, the system decouples into 1D elongated tubes and the particle exchange stops. To compute the population of atoms across the tubes, we solve the thermodynamics of an array of 1D tubes characterized by a unique temperature $T_\text{co}$ and with a local effective chemical potential $\mu_\text{1D}\!=\!\mu- \tfrac{1}{2}m\omega_z^2 z^2- \tfrac{1}{2}m\omega_y^2 y^2$, where $z$ and $y$ are the position of the 1D tubes in the transverse directions and $\omega_z$ and $\omega_y$ are the respective frequencies. We furthermore assume that the effective 1D interaction strength is the same as the one computed in the deeply 1D regime, at the end of the state-preparation protocol. The thermodynamics of the 1D tubes is determined by a thermodynamic Bethe ansatz \cite{takahashi2005thermodynamics} that we review in \secref{S_sec_GHD}.
\item As the lattice depth is increased from the dimensional crossover to the final deep 1D regime, the system evolves adiabatically, to a final temperature indicated as $T_\text{1D}$. We assume that $T_\text{1D}$ is constant across the tubes.
\item The two temperatures $T_\text{1D}$ and $T_\text{co}$ are used as fitting parameters and adjusted to ensure good agreement between theoretical data and experimental simulations on a reference dataset. We compare to the momentum distribution at $\ell\!=\!0$, as it has proven to be the most sensitive to temperature fluctuations.
\end{enumerate}

\noindent This theoretical modeling of the state preparation is a crude, yet effective approximation that describes well the experimental observations. There are several effects that are not captured by this approach, most prominently:
\begin{enumerate}
\item As the transverse trap frequency is increased from the dimensional crossover to the final 1D regime, the value of the 1D interaction strength changes \cite{Olshanii1998}. Accounting for this modification would require knowing the exact value of the transverse trapping at which the crossover happens.
\item The deeper the gas enters into the 1D regime, the better its dynamics is described by integrability, hindering thermalization. In each tube, the system at the end of the state-preparation protocol is arguably described by a GGE rather than a thermal Gibbs ensemble, even though we expect the difference to be small due to the good agreement between simulations and experimental data.
\item While we expect a unique temperature to characterize the system before the dimensional crossover, once the 1D tubes decouple, they can also acquire different temperatures. Therefore, a more precise modeling should keep into account variations of $T_\text{1D}$.
\end{enumerate}

\bigskip
\noindent Including the aforementioned effects in a theoretical modeling of the state preparation is a challenge for the future. A first improvement can be obtained following the adiabatic evolution from the dimensional crossover to the deep one-dimensional regime, and determine the final distribution of $T_\text{1D}$ through entropy conservation~\cite{Li2023}. Nonetheless, the simplified approach we consider is already sufficient to obtain a remarkable agreement with the experimental data. 
\bigskip

\noindent\textbf{The zero-temperature and strongly-repulsive approximation.---}
As a simplified model for the dimensional crossover, we can consider the limiting case in which the cross-over temperature is approximately zero, and in the regime where, at the dimensional crossover, the gas is already in the strongly repulsive TG regime. In this case, a simple analytical formula can be obtained.
Since the gas is in the fermionized regime, we can use the Thomas-Fermi approximation for fermions to describe the density profile within each tube, leading to 
\be
d_{y,z}(x)=\frac{1}{\pi}\sqrt{2m\mu\hbar^{-2}-(x/a_x^2)^2-(x/a_y^2)^2-(x/a_z^2)^2}\, ,\ee 
with $a_{x,y,z}=\sqrt{\hbar^2/(m\omega_{x,y,z})}$
the oscillator lengths of the three-dimensional trap. $\mu$ is the global chemical potential to be determined from the total number of atoms. Integrating along the $x$ direction, we obtain the number of atoms for a tube in position $(y,z)$ in the plane transverse to the 1D tubes as 
\be
N_{y,z}=\frac{1}{2a_x^2}\left[\tfrac{a_x^2 2m \mu}{\hbar^2}-\left(\tfrac{a_x^2}{a_y^2} y\right)^2-\left(\tfrac{a_x^2}{a_z^2} z\right)^2\right]\,.
\ee
We now consider the number of tubes $P(N)$ with $N$ atoms. The tubes in the y-z plane are arranged on a regular square grid with spacing $u=0.532 \mu\text{m}$.  We consider a continuum approximation in such a way $P(N)=\frac{1}{u^2}\int \dd y\dd z\, \delta\left(N_{y,z}-N\right)$. Notice that the total number of atoms $N_\text{atm}$ is given by $\int \dd N\, N P(N)=N_\text{atm}$.
The integral defining $P(N)$ can be easily computed, and the chemical potential expressed in terms of the total number of atoms, leading to the simple result
\be\label{eq_P_TF}
P(N)=\frac{2 N_\text{atm}}{N_\text{max}^2}\, \hspace{1pc}\text{for}\,\,\, N<N_\text{max}\, ,
\ee
and zero otherwise, where $N_\text{max}$ is determined as
\be\label{eq_Nmax}
N_\text{max}=\frac{u a_x}{a_y a_z}\sqrt{N_\text{atm}/\pi}\, .
\ee

\begin{figure}[t!]
    \centering
\includegraphics[width=\columnwidth]{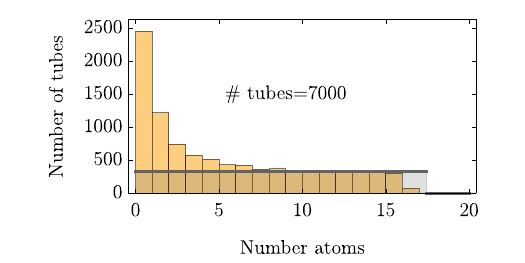}
    \caption{
    \textbf{Atom-number distribution across the 1D tubes.---} Using the method detailed in Note \ref{S_sec_prep}, we theoretically estimate the number of tubes with a given number of atoms after the loading protocol. The shaded black region corresponds to the approximation Eq.~\eqref{eq_P_TF}, where we get $N_\text{max}=17.45$. Since thermodynamics is solved in a local-density approximation, the number of atoms in each tube is not discrete, but it takes continuous values. We nevertheless show a histogram with unit spacing. To estimate the number of tubes that are effectively populated, we only consider the tubes holding more than one particle, resulting in $\sim7000$ occupied tubes. Results are shown for $T_\text{co}\!=\!1.5 \, \text{nK}$. Decreasing this value increases the number of atoms in each tube. The average number of atoms in the tubes is computed as $\langle N\rangle\!=\!\sum_i N^2_i/\sum_i N_i$, resulting in $\langle N\rangle\!=\!9.73$. Variations of $10\%$ in $T_\text{co}$ only slightly change the distribution (deviations are not shown), resulting in $\langle N\rangle\!=\!9.65$ and $\langle N\rangle\!=\!10.27$ for $T_\text{co}\!=\!1.65 \, \text{nK}$ and $T_\text{co}\!=\!1.35 \, \text{nK}$, respectively.}
    \label{Fig_SM_ntube}
\end{figure}

\bigskip
\noindent\textbf{Finite-temperature and finite interactions strength corrections.---}
Thermal fluctuations and a finite interaction strength modify the simple result derived above. To account for these effects, we solve the thermodynamics of 1D interacting Bose gases within the Thermodynamic Bethe Ansatz method (see Supplementary Note~\ref{S_sec_GHD}). As discussed in this Note, each tube has a local chemical potential renormalized by the effect of the shallow three-dimensional trap.

In Fig.~\ref{Fig_SM_ntube}, we show the theoretical number of atoms across the tubes with the best fitting parameters to the experimental data. The best fit is contrasted with the approximation from Eq.~\eqref{eq_P_TF}. The initial trap frequencies before loading the lattice are $\{\omega_x,\omega_y,\omega_z\}/2\pi\!=\!\{4,9,7\}\text{Hz}$, we assume that $\omega_x$ is also the longitudinal trap frequency at the dimensional crossover. We load $5\times 10^4$ atoms with a 3D scattering length $a_\text{3D}\!=\!500a_0$. The value of the 1D scattering length depends on the transverse trapping frequency \cite{Olshanii1998}, which is changing during the protocol, see also Note \ref{S_sec_GHD}. We approximate its value to be the same as the one at the end of the state preparation, resulting in $-2/a_\text{1D}\!=\!17.88 \mu\text{m}^{-1}$. The value of $T_\text{co}$ resulting in the momentum distribution that best compares with experimental data is $T_\text{co}\!=\!1.5 \, \text{nK}$.

\bigskip

\section{Supplementary note 6: Modeling of inhomogeneous FFS states via the Thomas-Fermi approximation}
\label{sec_TF}

For fermionic systems in a harmonic trap, the semiclassical Thomas-Fermi approximation provides a good description of the density profile and momentum distribution. Likewise, the bosonic FFS realized in our procedure are expected to be amenable to a similar approximation, provided that one accounts for the reduced occupancy $1/\ell$.  Within a single tube, we define a spatially-varying chemical potential $\mu(x)=\mu(0)-\frac{1}{2}m\omega_\parallel^2x^2$, which determines the local Fermi momentum $k_\text{F}(x)$ through the Fermi energy at $T$=0, $\frac{\hbar^2 k_\text{F}^2(x)}{2m}=\mu(x)$. The relation between the Fermi wave-vector and density $d(x)$ accounts for the reduced occupancy as $\ell \pi d(x)=k_\text{F}(x)$, and the global chemical potential $\mu(0)$ must be fixed from the normalization condition $N=\int \dd x\, d(x)$. The momentum distribution of the tube $n_N(k)$ is obtained integrating the local Fermi sea over the trap
\be
n_N(k)=\int \dd x\, \frac{1}{2\pi \ell}\theta(k_\text{F}(x)-|k|)\, ,
\ee
where $\theta(z)$ is the Heaviside step function such that $\theta(z>0)=1$ and $\theta(z<0)=0$. A simple computation leads to
\be
n_N(k)=\frac{a_\parallel}{N\pi \ell}\sqrt{2 N \ell-a^2_\parallel k^2}\, ,
\ee
where $a_\parallel$ is the oscillator length of the harmonic trap $a_\parallel=\sqrt{\hbar/(m\omega_\parallel)}$, and the normalization is $\int \dd k\, n_N(k)=1$.
The first-order correlator in the single tube $G_N^{(1)}(x)$ is then obtained through Fourier transform $G_N^{(1)}(x)=\int \dd k\, e^{ikx}n_N(k)$, resulting in
\be
G_N^{(1)}(x)=\frac{2}{(2N \ell)^{1/2}x a^{-1}_\parallel}J_{1}((2N \ell)^{1/2}x a^{-1}_\parallel)\, ,
\ee
with $J_1(z)$ being the Bessel function of the first kind. Since our experiment features several tubes with varying number of atoms $N$ distributed according to $P(N)$, with $\int \dd N\,P(N) N\!=\!N_\text{atm}$. The momentum distribution and $G^{(1)}(x)$ must be computed through a proper averaging
\begin{eqnarray}
\label{eq_SnN}n(k)&=&\frac{1}{N_\text{atm}}\int \dd N\, P(N) N n_N(k) \\
\label{eq_SGN}G^{(1)}(x)&=&\frac{1}{N_\text{atm}}\int \dd N\, P(N) N G^{(1)}_N(x)\, .
\end{eqnarray}

For a crude estimate, we consider the zero-temperature and strongly-repulsive approximation Eq.~\eqref{eq_P_TF}, leading to
\begin{eqnarray}
n(k)&\simeq&\frac{2 a_\parallel}{3\pi  \ell^2 N_\text{max}^2}[2\ell N_\text{max}-a_\parallel^2 k^2]^{3/2} \\
G^{(1)}(x)&\simeq&\frac{8}{(x a_\parallel^{-1})^2 2 \ell N_\text{max}}J_2(x a_\parallel^{-1} \sqrt{2 \ell N_\text{max}})\, .
\end{eqnarray}
Above, $N_\text{max}$ is determined from Eq.~\eqref{eq_Nmax}.
A more refined theoretical modeling requires a better characterization of the atom population across the tubes (see Supplementary Note \ref{S_sec_prep}). Thermal fluctuations in the initial state and other fine details can be modeled within GHD, see Supplementary Note \ref{S_sec_GHD}.

\bigskip

\section{Supplementary note 7: Theoretical modeling of the interaction cycle}
\label{S_sec_GHD}

In this Note we give an overview of the thermodynamics of the Lieb-Liniger model and the description of interaction strength cycles within GHD. More details are presented in Ref.~\cite{ThHolo}. Each 1D tube is microscopically well described by the Lieb-Liniger Hamiltonian with the addition of a trapping potential 

\be\label{eq_H_LL}
\hat{H} = -\sum_{i=1}^{N} \frac{\hbar^2}{2m}\frac{\partial^2}{\partial x_i^2} +\gone \sum_{ i < j } \delta(x_i - x_j)+\sum_i V(x_i),
\ee
where the potential $V(x)\!=\!\frac{1}{2}m\omega_{\parallel}^2 x^2$ describes the harmonic longitudinal confinement.
In the limit of a sufficiently smooth trapping potential and slow changes of the interaction strength $\gone$, GHD describes the system's evolution within a hydrodynamic framework. In a local-density approximation, the 1D system is assumed to be locally described by a Generalized Gibbs Ensemble (GGE), whose evolution is governed by the GHD equations. This section provides the rudiments of this approach and the necessary details to describe our experimental setting. The interested reader can further refer to Ref.~\cite{takahashi2005thermodynamics} for the thermodynamics on integrable systems, to Refs. \cite{Bastianello2022,Doyon2024} for pedagogical reviews on GHD, and to Ref.~\cite{ThHolo} for further details on the solution of the cycle within the GHD approach.

\bigskip

\noindent\textbf{Thermodynamic Bethe ansatz.---} We first discuss the thermodynamic description of integrable models within the framework of the thermodynamic Bethe ansatz (TBA) \cite{takahashi2005thermodynamics}, which lays the basis for the forthcoming hydrodynamic description. For the time being, we do not consider the trap and thus set $V(x)\!=\!0$: the effect of the trap will be restored later within a local-density approximation. A prominent peculiarity of integrable models is the existence of stable quasiparticles characterized by a quantum number $\lambda$, called ``rapidity", which generalizes to the interacting case the notion of a wave-vector of non-interacting particles.
To describe thermodynamics, in integrable models one introduces a rapidity density $\rho(\lambda)$, also known as root density, which plays the same role of the momentum distribution in non-interacting systems. Indeed, in the limit of vanishing interaction strength $\gone\to 0$, the rapidity distribution and momentum distribution coincide, but they differ as the interactions get stronger.
In principle, for attractive interactions $\gone<0$ the gas can feature bound states called Bethe strings: these compounds are excited when, for example, interactions are slowly driven from the repulsive to the attractive phase, passing through the non-interacting point \cite{horvath2025}. If Bethe strings are present, one needs to introduce several root densities, one for each species of bound state of $j$ particles. However, in the interactions strength cycle, Bethe strings' excitation is suppressed~\cite{Astrakharchik2005,Batchelor2005,Haller2009,ThHolo}. For simplicity, we discuss thermodynamics, and later GHD, in the absence of Bethe strings.
Each quasiparticle experiences a phase space renormalized by interactions with other excitations, leading to the concept of the total root density $\rho^t(\lambda)$ which, in the Lieb-Liniger model, is defined as
\be\label{eq_rhot}
\rho^t(\lambda)=\frac{1}{2\pi}-\int \frac{\dd\lambda'}{2\pi}\varphi(\lambda-\lambda')\rho(\lambda')\, ,
\ee
where $\varphi(\lambda)=-\frac{2m \gone/\hbar^2}{\lambda^2+(m \gone/\hbar^2)^2}$. The total root density describes the maximum density of states compatible with a fixed interaction strength. Notice that in the TG limit $\gone\to\infty$ one has $\rho^t(\lambda)=\frac{1}{2\pi}$ which coincides with the expectation for a free-fermion gas. Indeed, in the TG limit the interacting Bose gas fermionizes \cite{Paredes2004,Kinoshita2004}.
Together with $\rho(\lambda)$ and $\rho^t(\lambda)$, it is also convenient to introduce the occupancy or filling fraction $\vartheta(\lambda)=\rho(\lambda)/\rho^t(\lambda)$.
We can now consider a thermal Gibbs ensemble characterized by an inverse temperature $\beta=\frac{1}{K_\text{B} T_\text{1D}}$ and a chemical potential $\mu_{\text{1D}}$. Within the framework of thermodynamic Bethe ansatz, one can determine the free energy of the gas and, from its minimization, fix the rapidity density that describes thermal states. In this case, the root density, or equivalently the occupancy, must solve the non-linear integral equations
\be\label{eq_TBA}
\varepsilon(\lambda)=\beta (\epsilon(\lambda)-\mu_{\text{1D}})+\int \frac{\dd\lambda'}{2\pi}\varphi(\lambda-\lambda')\log(1+e^{-\varepsilon(\lambda')})\, ,
\ee
where one conveniently parametrizes the occupancy as $\vartheta(\lambda)=(1+e^{\varepsilon(\lambda)})^{-1}$ and $\epsilon(\lambda)=\frac{\hbar^2\lambda^2}{2m}$ is the kinetic energy of a single particle. These integral equations, as well as those defining the total root density \eqref{eq_rhot}, are straightforwardly discretized and numerically solved (see the end of this Supplementary Note). The TBA equation \eqref{eq_TBA} describes stable thermal states only in the repulsive phase: in the attractive regime, thermal states in the thermodynamic limit are unstable due to the presence of Bethe stings with arbitrary large binding energy \cite{McGuire1964,Koch2021} (not accounted for in Eq. \eqref{eq_TBA}). In our setup, only the initial state obtained after our state-preparation protocol (see \secref{S_sec_prep}) is described by a thermal state according to Eq. \eqref{eq_TBA}. The state is then pushed out of equilibrium during the cycle.

As a next step, we describe thermal states in the trapped Lieb-Liniger model \eqref{eq_H_LL} within a local density approximation, by promoting the rapidity distribution and occupancy to be spatially-dependent functions $\vartheta(\lambda)\to \vartheta_x(\lambda)$. For a fixed position $x$, the occupancy satisfies the TBA equation \eqref{eq_TBA} where the chemical potential is deformed by the longitudinal trap $\mu_{\text{1D}}\to \mu_{\text{1D}}-V(x)$.
Once the TBA equations have been solved, the local density of particles, $d(x)$ can be computed as $d(x)\!=\!\int \dd\lambda \rho_x(\lambda)$, resulting in a total number of particles $N\!=\!\int \dd x\, d(x)$.
Likewise the total internal energy $U\equiv\Big\langle -\sum_{i=1}^{N} \frac{\hbar^2}{2m}\frac{\partial^2}{\partial x_i^2} + \gone \sum_{ i < j } \delta(x_i - x_j)\Big\rangle$ and the potential energy $E_V\equiv \Big\langle \sum_i V(x_i) \Big\rangle$ are computable as
\be
U=\int \dd x\int \dd\lambda\, \epsilon(\lambda) \rho_x(\lambda)\, ,\,\,\, E_V=\int \dd x\int \dd\lambda\, V(x) \rho_x(\lambda)\, .
\ee
\bigskip

\noindent\textbf{Generalized hydrodynamics.---} So far, we discussed thermodynamics framed within TBA. 
Nonequilibrium protocols can instead be addressed by GHD, promoting the GGE to a time-dependent concept.
To this end, one introduces also a time-dependence in the root density $\rho_{x}(\lambda)\to \rho_{t,x}(\lambda)$. The dynamics is governed by the hydrodynamic equations \cite{Alvaredo2016,Bertini2016,Doyon2017,Bastianello2019}
\be\label{eq_ghd}
\partial_t \rho_{t,x}(\lambda)+\partial_x[v^\eff(\lambda)\rho_{t,x}(\lambda)]+\partial_\lambda[a^\eff(\lambda)\rho_{t,x}(\lambda)]\!=\!0\, .
\ee
Above, $v^\eff(\lambda)$ and $a^\eff(\lambda)$ are the effective velocity and acceleration, respectively. Both the velocity and acceleration are renormalized by the non-trivial scattering of particles and they explicitly depend on the state, making Eq.~\eqref{eq_ghd} a highly non-linear equation. More precisely, $v^\eff$ and $a^\eff$ are defined as
\be
v^\eff=\frac{(\partial_\lambda \epsilon)^\dr}{(\partial_\lambda p)^\dr}\, , \hspace{2pc}a^\eff=\frac{(f(\lambda))^\dr}{(\partial_\lambda p)^\dr}\, ,
\ee
where above $p(\lambda)\!=\!\hbar\lambda/m$ is the bare momentum and the bare force term is \cite{Doyon2017,Bastianello2019}
\be
f(\lambda) \!=\! -\partial_x V+\partial_t \gone\int \dd\lambda' \partial_{\gone}\Theta(\lambda-\lambda')\rho(\lambda')\, .
\ee
Above, $\partial_{\gone}\Theta(\lambda)$ is the derivative with respect to the interactions of the two-body scattering phase, which in Lieb-Liniger reads $\partial_{\gone}\Theta(\lambda)\!=\!2 \hbar^2 m\lambda/((m \gone)^2+(\hbar^2\lambda)^2)$.
Moreover, for an arbitrary function $\tau(\lambda)$ one defines the dressing operation $\tau(\lambda)\to \tau^\dr(\lambda)$ as the solution of the integral equation $\tau^\dr(\lambda)\!=\!\tau(\lambda)-\int \frac{\dd\lambda'}{2\pi}\varphi(\lambda-\lambda')\vartheta(\lambda')\tau^\dr(\lambda')$.

The GHD equation \eqref{eq_ghd} describes the evolution of the state within the repulsive $(\gone\!>\!0)$ and attractive $(\gone\!<\!0)$ branches of the interaction cycle. When crossing the TG-sTG transition and the non-interacting point $\gone\!=\!0$, proper boundary conditions to continue the root density in the next phase should be given: in both cases, $\rho_{t,x}(\lambda)$ is a continuous function, whereas the occupancy $\vartheta_{t,x}(\lambda)$ has a jump discontinuity crossing the non-interacting point \cite{ThHolo}. The GHD equations are numerically integrated by the standard method of characteristics \cite{Bastianello2019,Moller2020}.

\bigskip

\noindent\textbf{Computing the momentum distribution and $G^{(1)}(x)$.---} For general values of $\gone$, determining the momentum distribution associated with a given $\rho_{t,x}(\lambda)$ is challenging \cite{senese2025,ThHolo}. However, as previously mentioned, the root density converges to the momentum distribution in the limit $\gone\to 0$, which is also the regime in which the experimental momentum distribution is measured. This allows a direct comparison between the measurements and the GHD predictions.

\begin{figure}[b!]
    \centering
    \includegraphics[width=\columnwidth]{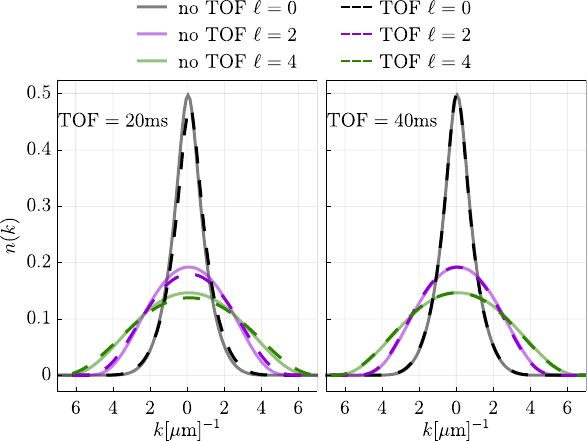}
    \caption{\textbf{Effect of a finite ToF on momentum distribution measurements.---}Theoretical comparison of the finite ToF momentum distribution (dashed lines) to the infinite ToF (solid line) result. The momentum is given as $k\!=\!p/\hbar$. Data are shown for $T_\text{1D}\!=\!5 \, \text{nK}$ and the average over the tubes' atom population is taken into account. Left panel: data for $\text{ToF}\!=\!20 \, \text{ms}$ are shown, small differences are still visible. Right panel: data for $\text{ToF}\!=\!40 \, \text{ms}$ (same value as for the experimental measurements in the main text) are shown, showing no appreciable deviation.}
    \label{Fig_SM_finite_TOF}
\end{figure}

In the simulations, we take into account finite-size and finite-ToF effects as follows (see also Ref.~\cite{horvath2025}). For $\gone\!=\!0$, right after the transverse confinement is removed, we can compute the local momentum density $n_x(p)$ in the 1D system from the rapidity distribution
$n_x(p)\equiv \hbar^{-1}\rho_x(\lambda)\Big|_{p=\hbar\lambda}$, the cloud is then expanded and imaged after a time-of-flight $t_\text{ToF}$, resulting in the 1D density profile $d_\text{ToF}(x)$
\be
d_\text{ToF}(x)=\int \dd x'\int \dd p' \,\delta\left(x'+\frac{p}{m}t_\text{ToF}-x\right)n_{x'}(p)\, .
\ee
Finally, we define the ToF momentum distribution as \be
n_\text{ToF}(p)=\frac{t_\text{ToF}}{m}d_\text{ToF}\left(\frac{p}{m}t_\text{ToF}\right),
\ee
which is ultimately compared to the experimental data, after averaging on the distribution of the number of atoms in the tubes, computed according to \secref{S_sec_prep}.
Notice that in the limit of infinite time-of-flight $n_\text{ToF}(p)$ approaches the momentum distribution integrated in the trap $\lim_{t_\text{ToF}\to\infty}n_\text{ToF}(p)=\int \dd x\, n_{x}(p)$. For finite ToF, corrections are visible. In Fig. \ref{Fig_SM_finite_TOF} we theoretically compare momentum distributions for different finite ToF with the infinite-ToF limit. While for $\text{ToF}\!=\!20 \, \text{ms}$ small deviations are visible, for $\text{ToF}\!=\!40 \,\text{ms}$ (i.e., the ToF used in the experiment) no appreciable differences are detected, indicating that the experimental data are indeed probing the momentum distribution with negligible corrections.
Finally, the theoretical $G^{(1)}(x)$ is computed through the Fourier transform of the momentum distribution, with the same procedure implemented on the experimental dataset.

\bigskip
\noindent \textbf{Simulating the full experimental protocol.---}
We here discuss the details of the experimental protocol and the determination of the physical parameters entering in the GHD equations. The 1D interaction strength $\gone$ is connected to the 1D scattering length $a_\text{1D}$ via $\gone\!=\!-2\hbar^2/(m a_\text{1D})$, then the 1D scattering length is determined by the 3D scattering length $a_\text{3D}$ as $a_\text{1D}\!=\!\frac{-a_\perp^2}{a_\text{3D}}(1-C a_\text{3D}/a_\perp)$ with $C\!=\!1.0326...$ and $a_\perp\!=\!\sqrt{\hbar/(m\omega_\perp)}$ \cite{Olshanii1998}. The experiment employs a fixed transverse frequency $\omega_\perp\!=\!2\pi\!\times\! 15\, \text{kHz}$ and $a_\text{3D}$ is tuned on the basis of overlapping Feshbach resonances~\cite{Haller2009} by varying the magnetic field $B$. The dependency is well approximated by
\be
a_\text{3D}\!=\!a_\text{bg}\prod_j \frac{B-Z_j}{B-R_j}\, ,
\ee
with experimentally determined parameters $a_\text{bg}\!=\!496 \, a_0$, $Z\!=\!\{17.1482,47.944,53,457,-90.68\}$ G, and $R\!=\!\{-12.2357,47.78,53,449,-126.23\}$ G.
During state preparation, the 1D tubes are prepared from a BEC holding $5\times 10^4$ atoms, with trap frequencies $\{\omega_x,\omega_y,\omega_z\}/2\pi\!=\! \{4,9,7\}\text{Hz}$\ and the magnetic field $B$ is set to $B\!=\!27.9\text{G}$. These parameters are used to estimate the atoms' population in each tube at the dimensional crossover by tuning $T_\text{co}$. After the dimensional crossover, while the system is entering in the 1D regime, the longitudinal trapping frequency is gently increased to $\omega_z\to 2\pi\times 18\,\text{Hz}$ and then kept fixed: this trapping frequency is used to run the thermodynamics of the 1D tubes and estimate $T_\text{1D}$.

\begin{figure*}[t!]
    \centering
\includegraphics[width=\textwidth]{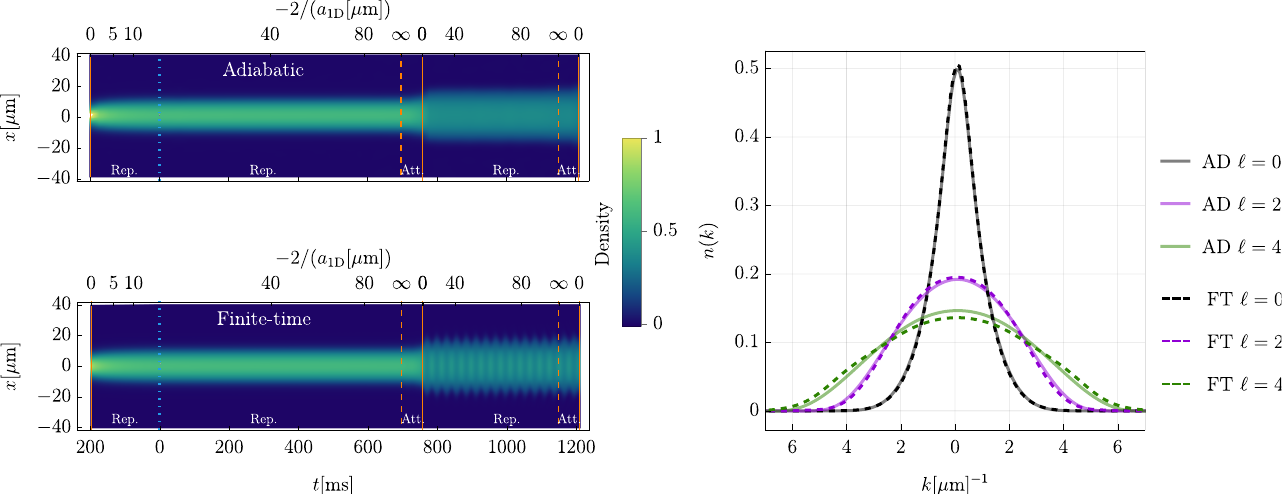}
    \caption{\textbf{Finite-time effects in the interactions strength cycle.---}We theoretically explore, within GHD, an adiabatic (AD) slow change in the interaction strength $\gone$ compared with the finite-time (FT) protocol implemented in the experiment. In the left panel, we compare the evolution of the density profile for a single representative tube with 10 atoms prepared with $T_\text{1D}=5\text{nK}$. The AD protocol (top panel) is compared with the FT protocol (bottom panel). On the bottom axis we show the evolution time of the protocol: starting from $t=0$ (blue dotted line) and moving on the right, the $\ell=2$ and $\ell=4$ states are reached. Moving on the left, $\ell=0$ is reached after 200ms of evolution. On the top axis, we show the interaction strengths in units $-2/a_\text{1D}$ corresponding to finite times by reporting some representative values. The TG points and zero crossing are highlighted with dashed and solid orange lines respectively. Regions with repulsive (Rep.) and attractive (Att.) interaction strengths are also highlighted. In the AD panel the time axis is not reported: to compare AD with FT, we show AD data at the interaction strength corresponding to the time evolution of the FT protocol. 
    In the right panel, we compare the theoretical momentum distribution for the AD (solid lines) and FT (dashed lines) protocol, showing data for $T_\text{1D}=5\text{nK}$ and keeping into account tubes' averaging. While for $\ell=0$ there is no appreciable difference, some small deviations are present for $\ell=2$ and $\ell=4$ due to the activation of the breathing mode (see main text).
    }
    \label{Fig_SM_finiteT}
\end{figure*}

We compare experimental data with fully adiabatic GHD simulations, namely, we consider an idealized and very slow ramp in the interaction strength $\gone$: the ramp rate is reduced until convergence is reached. 

However, the experiment implements a finite-time protocol. We now theoretically investigate finite-time protocols, and show that the corrections to the idealized case are negligible. In the experiment, after state preparation the magnetic field $B$ is changed accordingly to Fig. 2 of the main text, which we schematically summarize below

\begin{tikzcd}
& \arrow[d, ""] & & \\
27.9\text{G} \arrow[r, "500\text{ms}"] & 
40.85\text{G} \arrow[d, "200\text{ms}"]\arrow[r, "250\text{ms}"]&  
17.15\text{G} &  
 \\
 & 
47.20\text{G} \arrow[d, "quench"]\\
 & 
47.64\text{G}  \arrow[d, "250\text{ms}"]&  
 \\
 & 
49.01\text{G} \arrow[d, "quench"]&  
 \\
& \text{repeat} & & 
\end{tikzcd}

\bigskip

\noindent First, $B$ is ramped from $27.9$ G to $40.85$ G in $500$ ms. From this point, the cyclic operation starts. In the excited branches of the cycle, the measurement at $\gone\!=\!0$ is performed after the field is gently ramped to $17.15$ G ($\gone\simeq 0$) in 250 ms. Above, we give the experimental numbers for $B$: numerical simulations are done by rounding the magnetic fields to reach exactly $\gone\!=\!0$ at the non-interacting point, and to perform the TG-sTG transition at even stronger values of $|\gone|$, to ensure the continuity in the GHD equations.
Fig. \ref{Fig_SM_finiteT} compares the finite time (FT) protocol with the idealized adiabatic (AD) protocol. Due to the narrow CIR, the system features attractive interactions strengths ($\gone<0$) for much shorter times compared to the repulsive phase. This results in a relatively fast change of the interaction strength $\gone$ when higher excited states $\ell\!=\!2,4$ are accessed crossing $\gone\!=\!0$. This induces a small breathing mode visible in the simulations, which nonetheless has a small impact on the final momentum distribution.

\bigskip 
\noindent\textbf{Numerical discretization of GHD.---} We provide a short account of the numerical discretization that we use to solve the GHD equations. We use convenient units by measuring distances in $\mu\text{m}$, rapidities in units $\mu\text{m}^{-1}$, and the interaction strength is parametrized as $\tilde{c}\!=\!-2/(a_\text{1D}[\mu\text{m}])$, similarly to how it is done in Ref.~\cite{horvath2025}. Time is rescaled as $\tau\!=\!t/t_\text{eff}$, with $t_\text{eff}\!=\!\frac{1}{2} \frac{\hbar}{m u^2}$ and $u\!=\!1 \mu\text{m}$. First, the rapidity space is truncated $\lambda \in[-\Lambda,\Lambda]$ and discretized on a uniform grid $\{\lambda_i\}_{i=1}^{N_\lambda}$ with spacing $\dd \lambda \!=\! 2\Lambda/N_\lambda$. The real space is discretized similarly, within an interval $x\in[-X,X]$ using $N_x$ points. The rapidity distribution and the filling fractions become vectors in the discretized space $\rho_{t,x}(\lambda)\to \rho_{t,x_i}(\lambda_j)$. Integral equations are discretized accordingly, for example the integral appearing in the definition of $\rho^t$ in Eq.~\eqref{eq_rhot} becomes
\be
\int \frac{\dd\lambda'}{2\pi}\varphi(\lambda_j-\lambda')\rho(\lambda')\to \frac{1}{2\pi}\sum_{j'} \bar{\varphi}_{j,j'}\rho(\lambda_{j'})\, ,
\ee
where $\bar{\varphi}_{j,j'}\equiv \int_{\lambda_{j'}-\dd\lambda/2}^{\lambda_{j'}+\dd\lambda/2}\dd\lambda'\, \varphi(\lambda_j-\lambda')$ and the integral can be easily computed analytically. Using the exact integration of the kernel on the smooth integral rather than a middle-point approximation rule ensures that it remains well behaved for small $\gone$, as in this limit $\lim_{\gone\to 0}\varphi(\lambda)\!=\!-\text{sign}(\gone)\delta(\lambda)$. After discretization, the integral equations appearing in the TBA become matrix equations and are solved with standard libraries.
To propagate the time evolution predicted by GHD, we employ the method of characteristics \cite{Bastianello2019,Moller2020} (see also Ref.~\cite{ThHolo}) with a second-order algorithm. We observe that the finite-time protocol run by the experiment excites a small breathing of the atomic cloud not visible in the experiment, but that in the numerical simulations it causes a growing error that must be reduced by improving the discretization. We reach good convergence using $\Lambda\!=\! 8 [\mu\text{m}]^{-1}$, $X\!=\!40 \mu\text{m}$, $N_\lambda\!=\!N_x\!=\!200$, and $\dd\tau\!=\!5\times 10^{-4}$.

\bibliography{biblio_exp_new}